%% file: ms.tex
\documentclass[runningheads]{llncs}

\usepackage{amsmath,amssymb}
\usepackage{graphicx}
\usepackage{hyperref}

\usepackage{todonotes}
\usepackage{listings}
\usepackage{courier}
\usepackage{wrapfig}
\usepackage{ wasysym }
\usepackage{scalerel}

\input ztmpops
\input macros

\begin{document}
\title{RTAMT: Online Robustness Monitors from STL}
%
%
\author{Dejan Ni\v{c}kovi\'{c}\inst{1} \and
Tomoya Yamaguchi\inst{2}}
%
%
\institute{
AIT Austrian Institute of Technology,
Vienna, Austria
\and
Toyota Research Institute - North America, Ann Arbor, Michigan, U.S.\\
}
\maketitle              
\begin{abstract}
\input{abstract}
\end{abstract}
%
%
%
\section{Introduction}
\label{sec:intro}

\input{intro}


\subsubsection{Related Work}
\input{related}

\vspace{-15pt}
\section{RTAMT Design and Functionality}
\label{sec:rtamt}

\input{rtamt}

\input{ros}

\vspace{-15pt}
\section{Experiments and Use Scenario}
\label{Sec:Experiments}

\input{experiments}

\vspace{-15pt}
\section{Conclusions}
\label{Sec:Conclusion}

\input{conclusion}

\bibliography{cites}
\bibliographystyle{unsrt}

\newpage
\appendix



\section{Signal Temporal Logic}
\label{sec:stl}

\input{stl}

\input scenarios

\end{document}

%% file: ztmpops.tex
\newcommand\Dash{\scaleobj{0.9}{$\textendash$}}
\newcommand\Dia{\scaleobj{1.1}{\raisebox{-.4ex}
		{\rotatebox{45}{$\square$}}}}

\newcommand\BoxxB{\scaleobj{1.3}{\raisebox{-.05ex}{$\square$}}}

\newcommand\cir{\scaleobj{1.3}{\ocircle}}
\newcommand\dia{\Dia}
\newcommand\boc{\BoxxB}

\newcommand\cirM{\scaleobj{1.3}{\circleddash}}
\newcommand\diaM{\mbox{\ \ooalign{$\raisebox{.35ex}\Dash$\cr\hidewidth\raisebox{-.0ex}
			{\hspace{+0.2pt}$\Dia$}\hidewidth\cr}\ }}
\newcommand\boxM{\mbox{\ \ooalign{$\raisebox{.35ex}\Dash$\cr\hidewidth\raisebox{-.0ex}
			{\hspace{+0.2pt}{$\BoxxB$}}\hidewidth\cr}\ }}

\newcommand\until{\hskip 1.5pt\hbox{$\mathcal{U}$}\hskip 0.5pt}
\newcommand\since{\hskip 1.5pt\hbox{$\mathcal{S}$}\hskip 0.5pt}

\newcommand\Next{\cir}

\newcommand\Previous{\cirM}

\newcommand\G{\boc}
\newcommand\F{\dia}
\newcommand\opH{\boxM}
\newcommand\opP{\diaM}

\newcommand\Y{\Previous}

\newcommand\fildia[1]{\opP}
\newcommand\filcir[1]{\Y}
\newcommand\filboc[1]{\opH}

\newcommand\true{\emm{True}}
\newcommand\false{\emm{false}}

%% file: macros.tex
\newcommand{\stl}{STL}
\newcommand{\bfstl}{bfSTL}
\newcommand{\pstl}{pSTL}
\newcommand{\iostl}{IA-STL}

\newcommand{\signal}{w}

\newcommand{\f}{\varphi}

\newcommand{\R}{\mathbb{R}}
\newcommand{\N}{\mathbb{N}}

\newcommand{\rtamt}{\textsf{rtamt}}
\newcommand{\rosrtamt}{\textsf{rtamt4ros}}

\renewcommand{\true}{\textsf{true}}
\renewcommand{\false}{\textsf{false}}

\DeclareMathOperator\sign{sign}

%% file: abstract.tex
We present \textsf{rtamt}, an online monitoring library for Signal Temporal Logic (STL) 
and its interface-aware variant (IA-STL), providing 
both discrete- and dense-time interpretation of the logic. 
We also introduce $\rosrtamt$, a tool that integrates \textsf{rtamt} with Robotic Operating System (ROS), a 
common environment for developing robotic applications. We evaluate $\rtamt$ and $\rosrtamt$ on two 
robotic case studies.

%% file: intro.tex
Robotic applications are 
complex autonomous cyber-physical systems (CPS). 
Robotic Operating System (ROS)~\cite{ros} provides a meta-operating 
system that helps development of robotic applications. 
Verification remains a bottleneck, as existing techniques 
do not scale to this level of complexity, thus making static safety assurance 
a very costly, if not impossible, activity.
Run-time assurance (RTA) is an alternative approach for ensuring the safe operation of robotic CPS 
that cannot be statically verified. RTA allows the use of untrusted components in a system that implements a safe fallback mechanism for (1) detecting  
anomalies during real-time system operations and (2) invoking a recovery mechanism that brings the system back to its safe operation. 
Runtime verification (RV) provides a reliable and rigorous way for finding violations in system executions and consequently represents a viable solution 
for the monitoring RTA component. 

Formal specifications play an important role in RV and enable formulating system properties. Signal Temporal Logic (STL)~\cite{mn04} is 
a formal specification language used to describe CPS properties. It admits 
{\em robustness} semantics that measure how far is an observed behavior from satisfying/violating a 
specification. 

We introduce $\rtamt$\footnote{\url{https://github.com/nickovic/rtamt}}, an online STL monitoring library.  
$\rtamt$ supports standard STL and its {\em interface-aware} extension (IA-STL)~\cite{iastl} as 
specification languages. It provides automated generation of {\em online robustness} monitors from 
specifications under both {\em discrete} and {\em continuous} interpretation of time. 
%
We also present $\rosrtamt$\footnote{\url{https://github.com/nickovic/rtamt4ros}}, an extension that integrates $\rtamt$ to ROS,
thus enabling 
the use of specification-based RV methods in robotic applications.
We assess the library on two robotic applications. 

%

%% file: related.tex
Several tools support {\em offline} monitoring of STL with qualitative (AMT2.0~\cite{amt2}) and quantitative semantics
(S-TaLiRo~\cite{staliro} and Breach~\cite{breach}). Reelay~\cite{reelay} implements past Metric Temporal Logic (MTL) monitors 
over discrete and continuous time and with qualitative and quantitative semantics. 
RVROS~\cite{rosrv} is a specification-agnostic monitoring framework for improving safety and security of robots using ROS. To the best of 
our knowledge, $\rtamt$/$\rosrtamt$ is the only tool that implements online robustness STL monitors with both future and past operators and 
ROS support.

%% file: rtamt.tex

\begin{wrapfigure}{l}{0.55\textwidth}
\vspace{-20pt}
\centering
\scalebox{0.5}{ 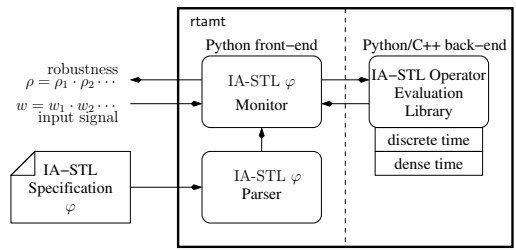 }
\caption{RTAMT architecture.}
\label{fig:arch}
\vspace{-20pt}
\end{wrapfigure}

The main functionality of \rtamt\ is the automatic generation of online robustness monitors from declarative specifications. 
Given an input signal in the form of a sequence of (time, value) pairs and a specification, $\rtamt$ computes at different points 
in time how robust is the observed signal to the specification, i.e. how far is it from satisfying or violating it.
The library consists of 3 major parts: 
(1) {\em specifications} expressed in a declarative specification language, (2) a {\em front-end} with an Application Programming Interface (API) 
to parse specifications and generate the monitor, and (3) a {\em back-end} that implements the actual evaluation algorithm used by the monitor.
The $\rtamt$ library uses a modular architecture depicted in Figure~\ref{fig:arch}. It uses ANTLR4 parser generator to 
translate textual (IA-)STL specifications into an abstract parse tree (APT) data structure used to build the actual monitor. 
The front-end implements the Application Programming Interface (API) and the pre-processing steps such as the translation 
of bounded-future (IA-)STL to past (IA-)STL in Python. The back-end implements the monitoring algorithms in Python 
(for discrete-time and dense-time interpretation) and C++ (for discrete-time interpretation). The library is compatible 
with both Python 2.7 and 3.7.



\noindent \textbf{Specification language}
in $\rtamt$ is STL with infinity-norm quantitative semantics~\cite{robust1}. 
The library supports four variants of the specification language -- standard STL and interface-aware STL~\cite{iastl} interpreted 
over discrete and dense time. IA-STL extends STL with an input/output signature of the variables and provides 
two additional semantic interpretations: (1) output robustness and (2) input vacuity. Output robustness measures 
robustness of output signals with respect to some fixed input. 
Input vacuity measures how vacuously is a specification satisfied with input signals only.
$\rtamt$ accepts as input {\em bounded-future} STL (\bfstl ) that restricts the use of the future temporal operators (eventually, always and until) 
to bounded intervals. See Appendix~\ref{sec:stl} for (IA-)STL definitions.

\noindent \textbf{Parsing and preprocessing}
follows a two-step procedure. The first step consists in translating 
the specification given in a textual form to an abstract parse tree (APT). The translation uses ANTLR4 
to generate a Python parser for the (IA-)STL grammar.
This translation is 
stll not suitable for online monitors -- the specification may have future temporal operator that would require clair-voyant 
monitoring capability. Hence, we implement the {\em pastification} procedure~\cite{bounded} that
translates the \bfstl\ formula $\phi$ into an {\em equi-satisfiable} 
past STL formula $\psi$, which uses only past temporal operators  and postpones
the formula evaluation from time index $t$, to the end of the (bounded) horizon $t+h$ where all the inputs necessary 
for computing the robustness degree are available (see Appendix~\ref{sec:pastification}).




\noindent \textbf{Monitoring}
consists of evaluating in online fashion the past $\stl$ specification according to its
quantitative semantics, interpreted in discrete or dense time 
\footnote{Due to pastification, 
$\rtamt$ only needs to evaluate past temporal operators.}.

\noindent \emph{Discrete-time monitors}  follow a time-triggered approach in which sensing of 
inputs and output generation are done at a periodic rate. 
This choice is motivated by~\cite{henzinger1992good}, which shows that by weakening/strengthening real-time specifications, 
discrete-time evaluation of properties preserves important properties of dense-time interpretation. This approach 
admits an upper bound on the use of computation resources.
$\rtamt$ implements two back-ends for STL monitors -- one in Python (for rapid prototyping) and 
one in C++ (for efficiency). $\rtamt$ uses Boost.Python library to integrate the Python front-end with 
the C++ backend. 

\noindent \emph{Dense-time monitors} follow an event-driven approach. Their implementation combines the incremental evaluation approach 
from~\cite{amt} with the optimal streaming algorithm to
compute the min and max of a numeric sequence over a sliding window from~\cite{robust2}. Unlike their
discrete-time counterparts, continuous-time monitors do not have bounds on memory requirements.

%% file: ov.pdf_t
\begin{picture}(0,0)%
\includegraphics{ov.pdf}%
\end{picture}%
\setlength{\unitlength}{3947sp}%
\begingroup\makeatletter\ifx\SetFigFont\undefined%
\gdef\SetFigFont#1#2#3#4#5{%
  \reset@font\fontsize{#1}{#2pt}%
  \fontfamily{#3}\fontseries{#4}\fontshape{#5}%
  \selectfont}%
\fi\endgroup%
\begin{picture}(6345,3066)(2689,-4594)
\put(3451,-4111){\makebox(0,0)[b]{\smash{{\SetFigFont{12}{14.4}{\rmdefault}{\mddefault}{\updefault}{\color[rgb]{0,0,0}$\varphi$}%
}}}}
\put(4051,-2536){\makebox(0,0)[rb]{\smash{{\SetFigFont{12}{14.4}{\rmdefault}{\mddefault}{\updefault}{\color[rgb]{0,0,0}$\rho = \rho_1 \cdot \rho_2 \cdots$}%
}}}}
\put(4051,-2386){\makebox(0,0)[rb]{\smash{{\SetFigFont{12}{14.4}{\rmdefault}{\mddefault}{\updefault}{\color[rgb]{0,0,0}robustness}%
}}}}
\put(5926,-3736){\makebox(0,0)[b]{\smash{{\SetFigFont{12}{14.4}{\rmdefault}{\mddefault}{\updefault}{\color[rgb]{0,0,0}IA-STL $\varphi$}%
}}}}
\put(5851,-2536){\makebox(0,0)[b]{\smash{{\SetFigFont{12}{14.4}{\rmdefault}{\mddefault}{\updefault}{\color[rgb]{0,0,0}IA-STL $\varphi$}%
}}}}
\put(4051,-2836){\makebox(0,0)[rb]{\smash{{\SetFigFont{12}{14.4}{\rmdefault}{\mddefault}{\updefault}{\color[rgb]{0,0,0}$w = w_1 \cdot w_2 \cdots$}%
}}}}
\put(4051,-2986){\makebox(0,0)[rb]{\smash{{\SetFigFont{12}{14.4}{\rmdefault}{\mddefault}{\updefault}{\color[rgb]{0,0,0}input signal}%
}}}}
\put(4951,-1786){\makebox(0,0)[lb]{\smash{{\SetFigFont{12}{14.4}{\rmdefault}{\mddefault}{\updefault}{\color[rgb]{0,0,0}$\rtamt$}%
}}}}
\end{picture}%

%% file: ros.tex
\begin{wrapfigure}{l}{0.5\textwidth}
\vspace{-20pt}
\centering
\scalebox{0.47}{ 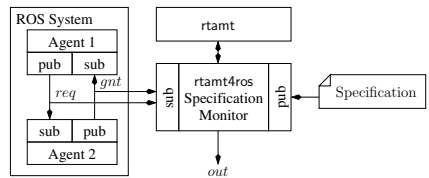 }
\caption{Integration of RTAMT to ROS.}
\label{fig:ros}
\vspace{-20pt}
\end{wrapfigure}

\noindent \textbf{Integration of RTAMT to ROS} ROS supports several messaging approaches, including the
{\em subscriber} and {\em publisher} pattern. 
A {\em publisher} categorizes a message into a class (called {\em topic} in ROS) and sends it withoutknowing who will read the message. 
A {\em subscriber} subscribes to a topic and receives its associated messages, without knowing who 
sent the message\footnote{Unless the publisher encodes its identity into the message itself.}. The messages are received and processed in 
\textsf{callback()} functions. Common ROS applications associate a \textsf{callback()} function per subscribed variable. 

$\rosrtamt$, illustrated in Figure~\ref{fig:ros}, integrates \rtamt\ into ROS using {\em rospy}. The integration is non-intrusive 
and provides the user with a generic and transparent monitoring solution for (IA-)STL specifications.  
The ROS system under observation is implemented with ROS nodes, which interact by publishing and 
receiving ROS messages on dedicated topics. To publish values of a variable $x$ of type $T$ on a topic $t$, ROS node associates 
$x$ and $T$ to $t$. Similarly, we declare in the STL specification variables that we want to monitor, declare their types and associate 
them to ROS subscription/publication topics using annotations. 
\noindent Variable names, their types and associated topics are specification-dependent. $\rosrtamt$ implements a 
{\em dynamic} subscription/publication mechanism that uses the concepts of {\em introspection} and {\em reflection} 
(the ability to passively infer the type of an object and actively change its properties at runtime). Given a (IA-)STL specification,
$\rosrtamt$ infers all the specification variables and  
dynamically creates their associated subscribers and publishers. The use of reflection 
allows us to associate a single \textsf{callback()} function to all specification variables, by passing the variable object as an argument 
to the function. 
We use the \textsf{callback()} function only to collect input data and the main ROS loop to make robustness monitor updates.


%% file: ros2.pdf_t
\begin{picture}(0,0)%
\includegraphics{ros2.pdf}%
\end{picture}%
\setlength{\unitlength}{3947sp}%
\begingroup\makeatletter\ifx\SetFigFont\undefined%
\gdef\SetFigFont#1#2#3#4#5{%
  \reset@font\fontsize{#1}{#2pt}%
  \fontfamily{#3}\fontseries{#4}\fontshape{#5}%
  \selectfont}%
\fi\endgroup%
\begin{picture}(5574,2326)(2164,-5675)
\put(5026,-4411){\makebox(0,0)[b]{\smash{{\SetFigFont{12}{14.4}{\rmdefault}{\mddefault}{\updefault}{\color[rgb]{0,0,0}$\rosrtamt$}%
}}}}
\put(3376,-4411){\makebox(0,0)[lb]{\smash{{\SetFigFont{12}{14.4}{\rmdefault}{\mddefault}{\updefault}{\color[rgb]{0,0,0}$gnt$}%
}}}}
\put(2776,-4561){\makebox(0,0)[lb]{\smash{{\SetFigFont{12}{14.4}{\rmdefault}{\mddefault}{\updefault}{\color[rgb]{0,0,0}$req$}%
}}}}
\put(4951,-3661){\makebox(0,0)[b]{\smash{{\SetFigFont{12}{14.4}{\rmdefault}{\mddefault}{\updefault}{\color[rgb]{0,0,0}$\textsf{rtamt}$}%
}}}}
\put(4951,-5611){\makebox(0,0)[b]{\smash{{\SetFigFont{12}{14.4}{\rmdefault}{\mddefault}{\updefault}{\color[rgb]{0,0,0}$out$}%
}}}}
\put(6526,-4561){\makebox(0,0)[lb]{\smash{{\SetFigFont{12}{14.4}{\rmdefault}{\mddefault}{\updefault}{\color[rgb]{0,0,0}Specification}%
}}}}
\end{picture}%

%% file: experiments.tex
We now present experiments performed using $\rtamt$ and $\rosrtamt$. 
We apply $\rtamt$ and $\rosrtamt$ on two ROS case studies: 
Simple Two Dimensional Simulator (STDR) and Toyota's Human Support Robot (HSR) platform~\cite{yamamoto2018development,yamamoto2019development}. 
We use the STDR example to show step-by-step usage of the $\rtamt$ and $\rosrtamt$ for online monitoring of robotic applications. 
We note that $\rtamt$ is versatile and could be used for instance for  
for offline monitoring and non-robotic applications (see Appendix~\ref{sec:scenarios}). We then evaluate the computation 
time requirements of the library.  
The experiments were performed on a Dell Latitude 7490 with an i7-8650U processor
and 16 GB of RAM, running Ubuntu 16.04 on a virtual machine.

\begin{figure}
\centering
\begin{minipage}{0.45\textwidth}
	\centering
	\includegraphics[width=\linewidth]{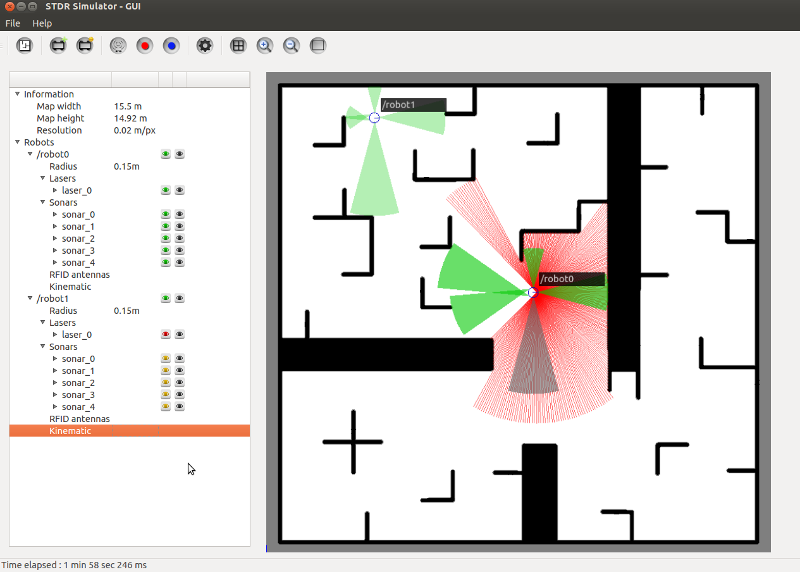}
	\caption{STDR simulator.}
  	\label{Fig:stdr}
\end{minipage}
\begin{minipage}{0.45\textwidth}
	\centering
	\includegraphics[width=\linewidth]{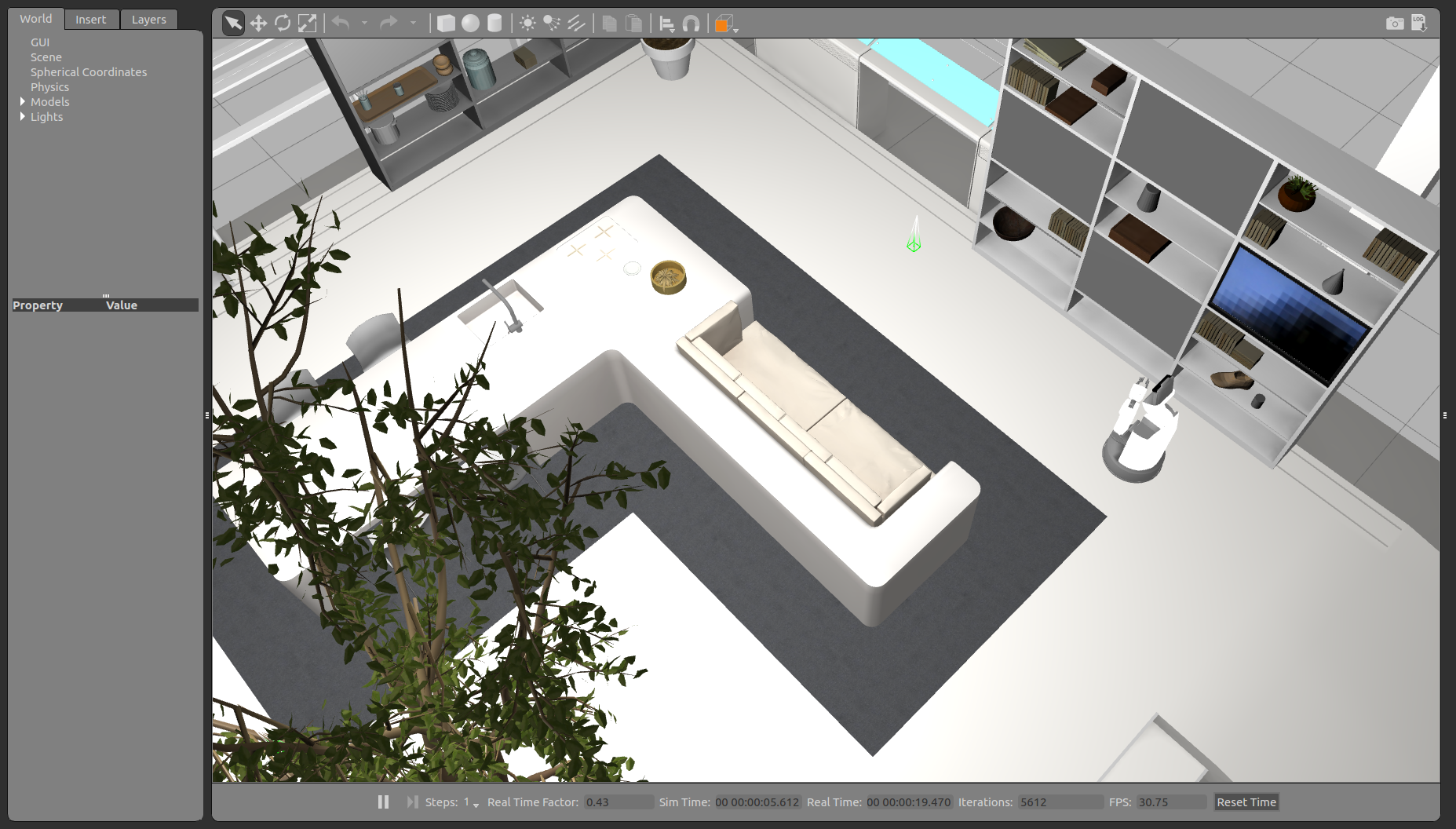}
	\caption{HSR service robotics application.}
  	\label{Fig:onlineSTL_HSR}
\end{minipage}
\vspace{-20pt}
\end{figure}

\noindent \textbf{Online monitoring of robotic applications:} 
STDR is a ROS-compliant environment for easy multi-robot 2D simulation (see Figure~\ref{Fig:stdr}). 
We use a simple robot controller with commands encoded as ROS \textsf{Twist} messages that expresses velocity in free space consisting 
of its linear and angular parts. 
The robot state is encoded as a ROS \textsf{Odometry} message that represents an estimate of the position (pose) and velocity (twist) in free space.
We than use the $\rosrtamt$ and $\rtamt$ to monitor its 
low-level requirement stating that every step in the command must be followed by the observed response. 
The specification \texttt{spec.stl} requires that at all times the distance between the linear velocity on the $x$ dimension of 
the command and the robot is smaller than $0.5$. The user first needs to import data types used in the specification (lines 1-3). 
Then, it declares variables used in specification, with their data type and (optionally) their input/output signature (lines 4, 6 and 8). 
Special comments in lines 5 and 7 are annotations that provide additional information about variables - in this case they associate 
variables to ROS topics. Finally, line 9 defines the IA-STL property.


\begin{lstlisting}
from geometry_msgs.msg import Twist
from nav_msgs.msg import Odometry
from rtamt_msgs.msg import FloatMessage
input Twist cmd
@ topic(cmd, robot0/cmd_vel)
output Odometry robot
@ topic(res, robot0/odom)
output FloatMessage out
out.value = always(abs(cmd.linear.x - robot.twist.twist.linear.x) <= 0.5)
\end{lstlisting}

To monitor the IA-STL specification \texttt{spec.stl} with $\rtamt$/$\rosrtamt$ 
, it suffices to run the following command in the ROS environment.
\begin{lstlisting}
roscore rtamt4ros ros_stl_monitor.py --stl spec.stl --period 100 --unit ms
\end{lstlisting}

\vspace{-20pt}

HSR is a robot with 8 degrees of freedom (DoF), combining 3 DoF of its mobile base, 4 DoF of the arm and 1 DoF of the torso lift (see Figure~\ref{Fig:onlineSTL_HSR}). 
The robot is equipped with ROS modules for localization, path planning and obstacle avoidance. We used this example to 
experiment with {\em system-level properties} in a multi-agent environment. We were interested in particular in 
monitoring the following requirements: (1) {\em no-collision} requirement stating that two robots are never closer than 
some $d_{\emph{min}}$ distance from each other, and (2) when robot 2 is closer than $d$ distance from robot 1, then 
robot 2 two goes in at most $T$ seconds within $d'$ distance of some location $L$. For this industrial application, we present 
an abstracted formalization of the above requirements.
\begin{lstlisting}
out1 = always (abs(rob1.pos - rob2.pos) < d)
out2 = abs(rob1.pos - rob2.pos) < d implies 
        eventually[0:T](rob1.pos - L) < d'
\end{lstlisting}
\noindent This experiment demonstrates the use of the library in a sophisticated ROS/Gazebo environment in an industrial case study. The 
addition of monitors is orthogonal to the development of the application and the monitors are non-intrusive.
\begin{wraptable}{l}{0.45\textwidth}
\centering
\vspace{-20pt}
	\begin{tabular}{|r | r | r|}
	\hline
	$k$ bound 	& C++ (s) 		& Python (s)		\\
	\hline
	100		& 0.00014		& 0.00023		\\	
	1k		& 0.0002		& 0.00085		\\
	10k		& 0.0008		& 0.029		\\
	100k		& 0.0047		& 0.31		\\
	1M 		& 0.046		& 72			\\
	\hline
	\end{tabular}
	\caption{Timing requirement per single monitor update.}
	\label{fig:cpp-pyth}
\vspace{-25pt}
\end{wraptable}

\noindent \textbf{Timing figures:} For online monitors, the most important quantitative measure is the computation time 
of a single monitoring update step. We compared the difference in timing requirements between 
the C++ and the Python implementation of the discrete-time monitoring algorithm. We used 
for the experiment the STL specification
\texttt{out = always[0:k] (a+b > -2)} where $k$ is the upper bound on the timing modality of the always operator that we
varied between $100$ and $1$ million. Table~\ref{fig:cpp-pyth} summarized the results of the 
experiment. The outcomes clearly demonstrate the efficiency of the C++ back-end, especially for large upper bounds in 
temporal modalities.

%% file: conclusion.tex
In this paper, we presented \rtamt\, a library for generating online monitors from declarative 
specifications and $\rosrtamt$, its ROS extension, demonstrating their usability and versatility two robotic case studies.

%% file: stl.tex
We consider Signal Temporal Logic ($\stl$) with both {\em past} and {\em future} operators with {\em quantitative} semantics and 
interpreted over discrete time. Let $X$ be a set of real-valued variables. A valuation $v~:~X \rightarrow \R$ For $x \in X$ maps a variable $x \in X$ to a real value. 
Given $Y \subseteq X$, we denote by $V(Y)$ the set of valuations over variables in $Y$. A {\em signal} $\signal$ over $X$ is a sequence 
$(t_0,v_0) \cdot (t_1,v_1) \cdots (t_n,v_n)$ of (time, valuation) pairs, where $t_0 = 0$ and $t_{i-1} < t_{i}$ for all $i$ in $[1,n]$. We note that 
for discrete-time interpretation $t_i = i$. For dense-time interpretation, we assume constant interpolation.  

The syntax of a \stl\ formula $\f$ over $X$ is defined by the following grammar:
$$
\f := f(Y) > 0~|~\neg \f~|~\f_1 \vee \f_2~|~\f_1 \until_I \f_2~|~\f_1 \since_I \f_2 
$$
\noindent where $Y \subseteq X$ and $I$ is of the form $[a,b]$ or $[a, \infty)$ such that 
$a$ and $b$ are in $\N$ and $0 \leq a \leq b$. 


The {\em quantitaive} semantics of a \stl\ formula with respect to a signal $w$ is described via the {\em robustness degree} $\rho(\f, \signal, t)$  indicating 
how far is the signal $\signal$ from satisfying or violating the specification $ \f$ at time index $t$.
$$
\begin{array}{lcl}
\rho(f(Y) > 0,w,t) & = & f(V(Y))[t] \\
\rho(\neg \f, w, t) & = & - \rho(\f, w, i) \\
\rho(\f_1 \vee \f_2, w, t) & = & \max(\rho(\f_1,w,t), \rho(\f_2,w,t))\\
\rho(\f_1 \since_I \f_2,w,t) & = & \max_{t' \in ((t \ominus I)  \cap \R_{\geq 0})}(\min(\rho(\f_2,w,t'), \min_{ t'' \in (t', t)} \rho(\f_1, w,t'')))\\
\rho(\f_1 \until_I \f_2,w,t) & = & \max_{t' \in (t \oplus I)}(\min(\rho(\f_2,w,t'), \min_{ t'' \in (t, t')} \rho(\f_1, w,t'')))\\
\end{array}
$$

From this basic definition of \stl\ we use standard rules to derive the other Boolean and temporal operators:
$$
\begin{array}{llclllcl}
\text{true} & \true & \equiv & \f \vee \f'  &
\text{false} & \false & \equiv & \neg \true \\
\text{conjunction} & \f_1 \wedge f_2 & \equiv & \neg(\neg \f_1 \vee \neg \f_2)) &
\text{eventually} & \F_I \f & \equiv & \true \until_I \f \\
\text{always} & \G_I \f & \equiv & \neg \F_I \neg \f &
\text{once} & \opP_I \f &  \equiv & \true \since_I \f \\
\text{historically} & \opH_I \f & \equiv & \neg \opP_I \ne \f &
\text{next} & \Next \f & \equiv & \false \until_{(0,\infty)} \f \\
\text{previous} & \Previous \f & \equiv & \false \since_{(0,\infty)} \f &
\text{rise} & \uparrow \f & \equiv & \Previous \neg \f \wedge \f \\
\text{fall} & \downarrow \f & \equiv & \Previous \f \wedge \neg \f &&&&\\
\end{array}
$$

We consider in this paper two subsets of \stl\:
\begin{itemize}
\item {\em Past} Signal Temporal Logic (\pstl ) that forbids the use of $\until_I$, and
\item {\em Bounded-Future} Signal Temporal Logic (\bfstl ) that restricts the use of $\until_I$ to intervals $[a,b]$ with both $a$ and $b$ in $\R_{\geq 0}$.
\end{itemize}

\subsection{Interface-aware Signal Temporal Logic}
\label{sec:ia-stl}
In this section, we consider \stl\ with input/output signature. Given a set $X$ of real-valued variables, 
let $U,V \subseteq X$ be two subsets of $X$ such that $U \cap V = \emptyset$. Let $\f$ be an STL formula and $w$ a signal trace.
We define the $U$-robustness relative to $V$, denoted $\rho_U^V(\f,w,t)$ by induction, where the only difference from the 
definition of  $\rho(\f,w,t)$ is the case on numeric predicates:
\begin{align*}
\rho_U^V(f(Y) > 0, w, t) &= 
	\begin{cases}
		0 &\text{if } Y \not \subseteq U \cup V \\
		f(w_Y[t]) &\text{else if } Y \not \subseteq V \\
		\sign(f(w_Y[t])) \cdot \infty &\text{otherwise}
	\end{cases}\\
\end{align*} 
where for all $a \in \R$, $\sign(a) \cdot \infty = +\infty$ if $a > 0$, $-\infty$ otherwise.

An {\em interface-aware signal temporal logic} (\iostl\ ) specification over $X$ is a tuple $(X_I, X_O, \f)$, where $X_I, X_O \subseteq X$, 
$X_I \cap X_O = \emptyset$ and $\f$ is a STL formula. We define two different notions of relative reobustness for \iostl\ specifications: 
(1) {\em output robustness}, denoted $\mu$, as the $X_O$-robustness relative to $X\backslash X_O$, and (2)
{\em input vacuity}, denoted $\nu$, the $X_I$-robustness relative to $\emptyset$.

\section{From Bounded-Future \stl\ to Past \pstl\ }
\label{sec:pastification}

Monitoring specifications with future temporal formulas is challenging because the evaluation at time index $t$ may depend on the observed 
inputs at some time indices $t' > t$. \bfstl\ specifications have a bounded future horizon $h$ that can be syntactically computed from 
the formula structure. For such specifications, the online monitoring challenge can be addressed by 
postponing the formula evaluation from time index $t$, to the end of the (bounded) horizon $t+h$ where all the inputs necessary 
for computing the robustness degree are available. In this section, we briefly sketch this procedure, called {\em pastification}.

We first define the {\em temporal depth} $H(\f)$ of $\f$ as the syntax-dependent upper bound on the actual depth of the specification that is 
inductively computed as follows:
$$
\begin{array}{lcl}
H(f(Y) \geq 0) & = & 0 \\
H(\neg \varphi) & = & H(\varphi) \\
H(\varphi_1 \vee \varphi_2) & = & \max \{ H(\varphi_1), H(\varphi_2) \} \\
H(\Next \varphi) & = & H(\varphi) + 1\\
H(\varphi_1 \until_I \varphi_2) & = & \textsf{end}(I) + \max \{ H(\varphi_1)-1, H(\varphi_2) \} \\
\end{array}
$$ 

The {\em pastification} operation $\Pi$ on the $\stl$ formula $\varphi$ with past and bounded future and 
its bounded horizon $d = H(\varphi)$ is defined recursively as follows:
$$
\begin{array}{lcl}
\Pi(p, d) & = & \opP_{\{d\}} p \\
\Pi(\neg \varphi, d) & = & \neg \Pi(\varphi, d) \\
\Pi(\varphi_1 \vee \varphi_2, d) & = & \Pi(\varphi_1, d) \vee \Pi(\varphi_2, d) \\
\Pi(\Next \varphi, d) & = & \Pi(\varphi, d-1) \\
\Pi(\F_{[a,b]} \varphi, d) & = & \opP_{[0,b-a]} \Pi(\varphi, d-b) \\
\Pi (\varphi_1 \until_{[a,b]} \varphi_2, d) & \leftrightarrow &  
\bigvee_{i = 1}^{z} \Pi (\varphi^i_1 \until_{[a,b]} \varphi_2, d) \\
&\leftrightarrow& \bigvee_{i = 1}^{z} \Pi(\Next \varphi^i_1 \wedge \\ 
&&\F_{[a-1, b-1]} 
(\varphi^i_1 \wedge \Next \varphi_{2}), d) \\
\end{array}
$$

Formally, we say 
that for an arbitrary $\bfstl$ formula $\f$, signal $w$ and time index $t \in \N$, $\rho(\f,w,t) = \rho(\Pi(\f),w,h(\f))$.

%% file: scenarios.tex
\section{Other Use Scenarios}
\label{sec:scenarios}

In this paper, we focused on the use of $\rtamt$ combined with $\rosrtamt$ in the context of robotic applications 
using ROS. However, $\rtamt$ is a much more versatile library and can be used in many other (non-robotic) 
applications. We illustrate the versatility of $\rtamt$ with two additional applications: (1) general offline anaysis of 
time series, and (2) model-based development of CPS applications with MATLAB/Simulink.

\subsection{Offline Analysis of Time Series with $\rtamt$}

\begin{figure}
\centering
\scalebox{0.6}{ 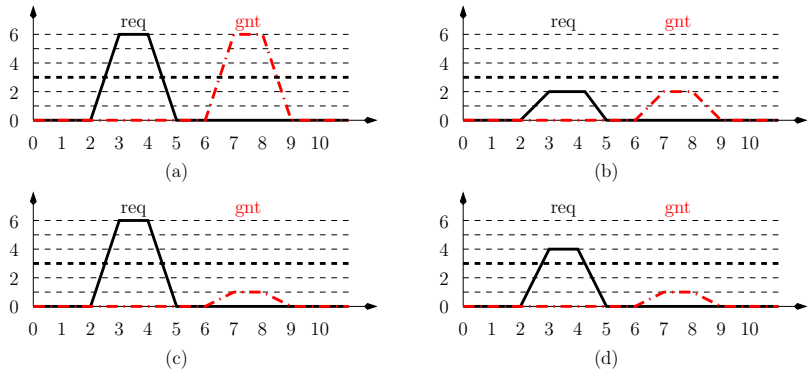 }
\caption{Examples of request-grant signals.}
\label{fig:example}
\end{figure}

We use a simplified real-valued request-grant protocol to illustrate (1) interface-aware STL semantics, (2) programmatic interface to $\rtamt$ and 
(3) offline monitoring capability of $\rtamt$. 
We say that a 
request is issued whenever the value of the signal \textsf{req} is greater or equal to $3$. Similarly, the grant is issued whenever the value of the signal 
\textsf{gnt} is greater or equalt to $3$. The specification requires that every request is eventually followed within $5$ time units by a grant. Figure~\ref{fig:example} 
illustrates $4$ different signals: (a) satisfies the requirement, (b) vacuously satisfies the requirement (request is never issued), (c) and (d) violate the requirement. 
We demonstrate the discrete-time offline monitoring by iterating over the input sequence and applying the online monitor in each iteration and collecting the output.

\iostl\ specification of the request-grant requirement 
from Section~\ref{sec:intro} is the formula $(\{ req \}, \{ gnt \}, \f)$, where $\f \equiv \G(req \geq 3 \to \F_{[0,5]} gnt \geq 3)$.
The text below expresses $\f$ using the \rtamt\ STL grammar. 
\begin{lstlisting}
import rtamt
...

data = read_csv('req-gnt.csv')

spec = rtamt.STLIOSpecification(is_pure_python=True)
spec.name = 'IA-STL Request-Grant Specification'
spec.declare_var('req', 'float')
spec.declare_var('gnt', 'float')
spec.declare_var('out', 'float')
spec.set_var_io_type('req', 'input')
spec.set_var_io_type('gnt', 'output')
spec.spec = 'out = always((req>=3) implies (eventually[0:5](gnt>=3)))'
spec.iosem = 'output-robustness'
try:
    spec.parse()
except rtamt.STLParseException as err:
    print('STL Parse Exception: {}'.format(err))
    sys.exit()

for i in range(len(data1[' gnt'])):
    rob = spec.update(i, [('req', data1[' req'][i][1]), ('gnt', data1[' gnt'][i][1])])

\end{lstlisting}

We first note that we instantiate the class \textsf{STLIOSpecification} to create an IA-STL monitor. We set the argument 
\textsf{is\_pure\_python} to \textsf{True} if we want to create a pure Python monitor. This flag is set to \textsf{False} for 
monitors using the C++ back-end. We observe that we use the method \textsf{set\_var\_io\_type(var\_name, var\_io\_type)} 
to declare a variable as input or output. We use the attribute \textsf{iosem} to define the IA-STL semantics that we want to use -- 
the available options are \textsf{standard}, \textsf{output-robustness} and \textsf{input-vacuity}.

Consider signals from Figure~\ref{fig:example} 
and the request-grant specification.  
Standard robustness $\rho(\f,w_4)$ is $-1$, while output robustness $\mu(\f,w_4)$ is $-2$ -- standard robustness gives the same treatment to input and output signals 
resulting the input dominating robustness computation. Standard robustness measures in this case how good is the input, while output measures how bad is the 
grant. We also have that $\mu(\f,w_2) = +\infty$ while $\nu(\f,w_2) = 1$. Output robustness value tells us that the input 
is vacuous, while the input vacuity measures how vacuous it is.   

The dense-time monitor for the same input signal and specification can be invoked by instantiating the \textsf{STLIOCTSpecification} instead of 
\textsf{STLIOSpecification}. We note that the signature of the \textsf{update} method is different for discrete-time and dense-time monitors:

\begin{tabular}{ll}
Discrete-time & \textsf{update(time\_index, [ (var\_name, value) ... ])} \\
Dense-time & \textsf{update([[var\_name, [[time, value] ...] ...])} \\
\end{tabular}

\subsection{Model-based design of CPS with $\rtamt$}

We now introduce a case study that we use as a running example to illustrate 
our approach step by step.  We consider the Aircraft Elevator Control System~\cite{mosterman-fdir} 
to illustrate model-based development of a 
Fault Detection, Isolation and Recovery (FDIR) application for a redundant actuator 
control system. 

\begin{figure*}
\centering
\scalebox{0.65}{ 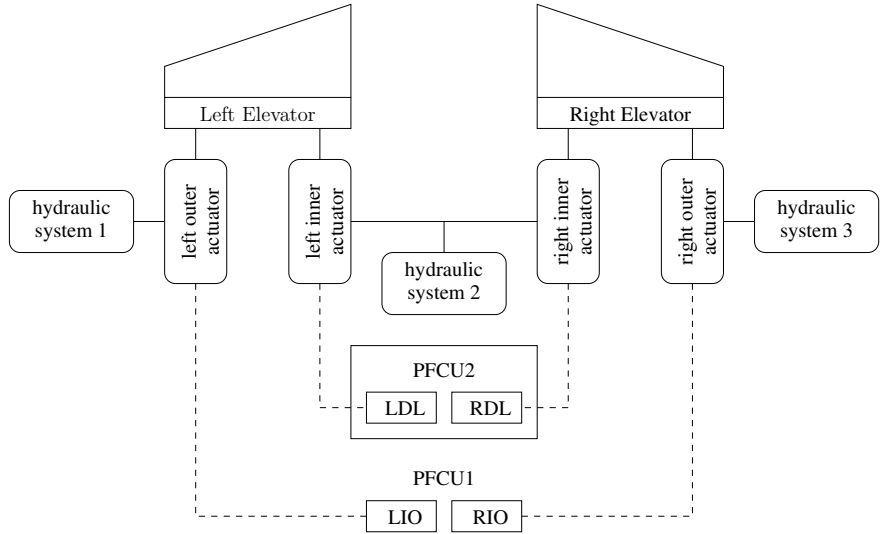 }
\caption{Aircraft Elevator Control System~\cite{mosterman-fdir}.}
\label{fig:aircraft}
\end{figure*}

Figure~\ref{fig:aircraft}  shows the architecture of an aircraft elevator control system with redundancy, with one elevator on the left and one on the right side. Each elevator is equipped with two hydraulic actuators. Both actuators can position the elevator, but only one shall be active at any point in time. There are three different hydraulic systems that drive the four actuators. The left (LIO) and right (RIO) outer actuators are controlled by a Primary Flight Control Unit (PFCU1) with a sophisticated input/output control law.  If a failure occurs, a less sophisticated Direct-Link (DL/PFCU2) control law with reduced functionality takes over to handle the left (LDL) and right (RDL) inner actuators. The system uses  state machines to coordinate the redundancy and assure its continual fail-operational activity. 

This model has one input variable, the input Pilot Command, and two output variables, the position of left and right actuators, as measured by the sensors. This is a complex model that could be extremely hard to analyze in case of failure. In fact, the model has $426$ signals, from which $361$ are internal variables that are instrumented ($279$ real-valued, $62$ Boolean and $20$ enumerated - state machine - variables) and any of them, or even a combination of them, might be responsible for an observed failure.

When the system behaves correctly, the intended position of the aircraft required by the pilot must be achieved within a predetermined time limit and with a certain accuracy. This can be captured with several requirements. One of them says that whenever Pilot Command $\emph{cmd}$ goes above a threshold $m$, the actuator position measured by the sensor must stabilize (become at most $n$ units away from the command signal) within 
$T+t$ time units. This requirement is formalized in STL with the following specification:
\begin{lstlisting}
out = always((cmd >= m) -> eventually[0:T]always[0,t](abs(cmd - pos) < n)
\end{lstlisting}

\begin{figure}[h]
\begin{center}
  \begin{minipage}[b]{0.48\columnwidth}
    \includegraphics[width=\columnwidth]{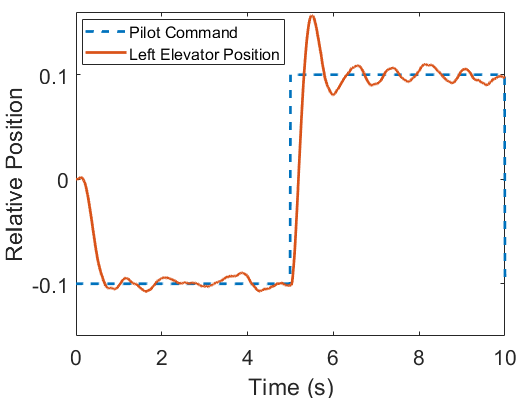}
    \caption{Expected behavior of AECS.}
    \label{fig:expected}
  \end{minipage}
  \hspace{0.1cm}
  \begin{minipage}[b]{0.48\columnwidth}
    \includegraphics[width=\columnwidth]{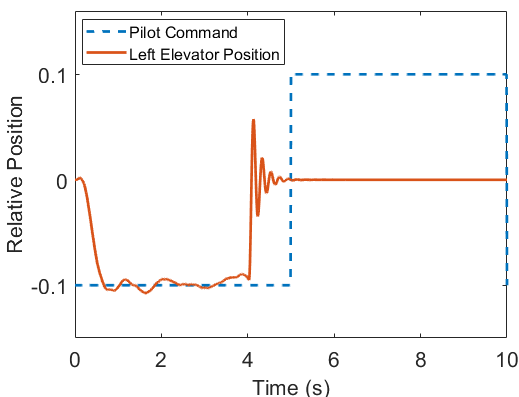}
    \caption{Failure of the AECS.}
    \label{fig:faulty}
  \end{minipage}  
  \end{center}
\vspace{-20pt}
\end{figure}
Figures~\ref{fig:expected} and~\ref{fig:faulty} show the correct and faulty behavior of the system, respectively. The control system clearly stops following the reference signal after $4$ seconds.

\begin{figure}
\centering
\begin{minipage}{0.45\textwidth}
	\centering
	\includegraphics[width=\linewidth]{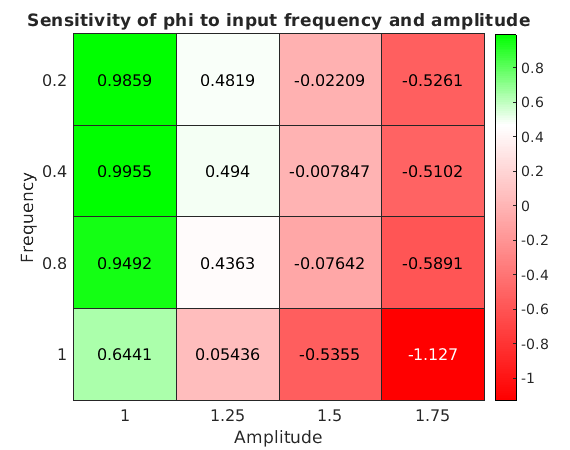}
	\caption{Sensitivity analysis of AECS.}
  	\label{fig:sa}
\end{minipage}
\begin{minipage}{0.45\textwidth}
	\centering
	\includegraphics[width=\linewidth]{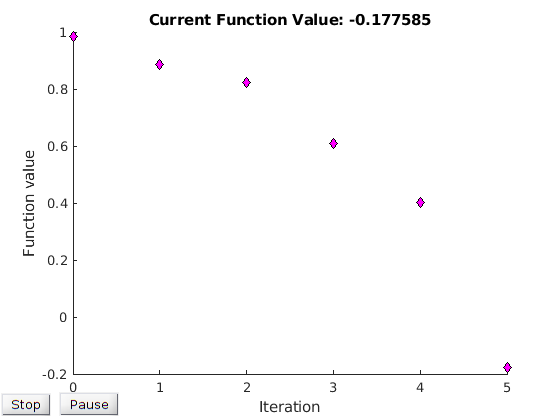}
	\caption{Falsification testing of AECS.}
  	\label{fig:ft}
\end{minipage}
\end{figure}

We use $\rtamt$ to do {\em sensitivity analysis} and {\em falsification testing}, two common approaches 
supported by tools like Breach and S-TaLiRo. 
Sensitivity of the model robustness to its \stl\ requirements is studied by uniformly varying input parameters, simulating
the model for each combination and monitoring the
simulation outcomes against the requirements. 
Figure~\ref{fig:sa} shows a heat-map visualizing the outcomes. 

Given an \stl\ specification and the
system model, falsification testing aims to identify an input signal that results
in the violation of the requirement. It does so by solving a search
problem over input parameter variables to identify a trace that violates the specification.
The outcome is shown in Figure~\ref{fig:ft}.

Finally, we integrate $\rtamt$ to MATLAB Simulink to 
evaluate specifications during simulation in an online fashion. We create a custom online monitoring Simulink block 
that integrates $\rtamt$ to a Level-2 MATLAB S-Function. At every simulation step, the monitoring block 
reads the relevant Simulink signal values and computes 
robustness for that simulation step. The robustness value is given back to the simulation environment. 
This approach has the advantage of being able to stop the simulation upon detection of property violation. 
This can save significant simulation time, especially when property violations are detected early in the simulation.

\section{Reproducing Examples}
\label{sec:rep}

All the examples shown in the paper and its Appendix, except the proprietary Toyota HSR robotic application, 
can be found on the repositories, together with instructions on how to run them.

\begin{tabular}{| l | l | l | l |}
\hline
Example & Repository & Dependencies & Directory \\
\hline
STDR & rtamt4ros & ROS, STDR & stdr\_control \\
Timing & rtamt & none & examples/timing \\
Offline req-gnt & rtamt & none & examples/offline\_monitors \\
Sensitivity analysis & rtamt & MATLAB/Simulink & examples/sensitivity\_analysis\_simulink \\
Falsification testing & rtamt & MATLAB/Simulink & examples/falsification\_testing\_simulink \\
Online Simulink & rtamt & MATLAB/Simulink & examples/online\_monitos\_matlab\_simulink \\
\hline
\end{tabular}

%% file: example.pdf_t
\begin{picture}(0,0)%
\includegraphics{example.pdf}%
\end{picture}%
\setlength{\unitlength}{3947sp}%
\begingroup\makeatletter\ifx\SetFigFont\undefined%
\gdef\SetFigFont#1#2#3#4#5{%
  \reset@font\fontsize{#1}{#2pt}%
  \fontfamily{#3}\fontseries{#4}\fontshape{#5}%
  \selectfont}%
\fi\endgroup%
\begin{picture}(8277,3826)(1036,-6125)
\put(7801,-3811){\makebox(0,0)[b]{\smash{{\SetFigFont{12}{14.4}{\rmdefault}{\mddefault}{\updefault}{\color[rgb]{0,0,0}$7$}%
}}}}
\put(6001,-3811){\makebox(0,0)[b]{\smash{{\SetFigFont{12}{14.4}{\rmdefault}{\mddefault}{\updefault}{\color[rgb]{0,0,0}$1$}%
}}}}
\put(6301,-3811){\makebox(0,0)[b]{\smash{{\SetFigFont{12}{14.4}{\rmdefault}{\mddefault}{\updefault}{\color[rgb]{0,0,0}$2$}%
}}}}
\put(6601,-3811){\makebox(0,0)[b]{\smash{{\SetFigFont{12}{14.4}{\rmdefault}{\mddefault}{\updefault}{\color[rgb]{0,0,0}$3$}%
}}}}
\put(6901,-3811){\makebox(0,0)[b]{\smash{{\SetFigFont{12}{14.4}{\rmdefault}{\mddefault}{\updefault}{\color[rgb]{0,0,0}$4$}%
}}}}
\put(7201,-3811){\makebox(0,0)[b]{\smash{{\SetFigFont{12}{14.4}{\rmdefault}{\mddefault}{\updefault}{\color[rgb]{0,0,0}$5$}%
}}}}
\put(7501,-3811){\makebox(0,0)[b]{\smash{{\SetFigFont{12}{14.4}{\rmdefault}{\mddefault}{\updefault}{\color[rgb]{0,0,0}$6$}%
}}}}
\put(8101,-3811){\makebox(0,0)[b]{\smash{{\SetFigFont{12}{14.4}{\rmdefault}{\mddefault}{\updefault}{\color[rgb]{0,0,0}$8$}%
}}}}
\put(8401,-3811){\makebox(0,0)[b]{\smash{{\SetFigFont{12}{14.4}{\rmdefault}{\mddefault}{\updefault}{\color[rgb]{0,0,0}$9$}%
}}}}
\put(8701,-3811){\makebox(0,0)[b]{\smash{{\SetFigFont{12}{14.4}{\rmdefault}{\mddefault}{\updefault}{\color[rgb]{0,0,0}$10$}%
}}}}
\put(5551,-3586){\makebox(0,0)[rb]{\smash{{\SetFigFont{12}{14.4}{\rmdefault}{\mddefault}{\updefault}{\color[rgb]{0,0,0}$0$}%
}}}}
\put(1201,-3811){\makebox(0,0)[b]{\smash{{\SetFigFont{12}{14.4}{\rmdefault}{\mddefault}{\updefault}{\color[rgb]{0,0,0}$0$}%
}}}}
\put(1501,-3811){\makebox(0,0)[b]{\smash{{\SetFigFont{12}{14.4}{\rmdefault}{\mddefault}{\updefault}{\color[rgb]{0,0,0}$1$}%
}}}}
\put(1801,-3811){\makebox(0,0)[b]{\smash{{\SetFigFont{12}{14.4}{\rmdefault}{\mddefault}{\updefault}{\color[rgb]{0,0,0}$2$}%
}}}}
\put(2101,-3811){\makebox(0,0)[b]{\smash{{\SetFigFont{12}{14.4}{\rmdefault}{\mddefault}{\updefault}{\color[rgb]{0,0,0}$3$}%
}}}}
\put(2401,-3811){\makebox(0,0)[b]{\smash{{\SetFigFont{12}{14.4}{\rmdefault}{\mddefault}{\updefault}{\color[rgb]{0,0,0}$4$}%
}}}}
\put(2701,-3811){\makebox(0,0)[b]{\smash{{\SetFigFont{12}{14.4}{\rmdefault}{\mddefault}{\updefault}{\color[rgb]{0,0,0}$5$}%
}}}}
\put(3001,-3811){\makebox(0,0)[b]{\smash{{\SetFigFont{12}{14.4}{\rmdefault}{\mddefault}{\updefault}{\color[rgb]{0,0,0}$6$}%
}}}}
\put(3301,-3811){\makebox(0,0)[b]{\smash{{\SetFigFont{12}{14.4}{\rmdefault}{\mddefault}{\updefault}{\color[rgb]{0,0,0}$7$}%
}}}}
\put(3601,-3811){\makebox(0,0)[b]{\smash{{\SetFigFont{12}{14.4}{\rmdefault}{\mddefault}{\updefault}{\color[rgb]{0,0,0}$8$}%
}}}}
\put(3901,-3811){\makebox(0,0)[b]{\smash{{\SetFigFont{12}{14.4}{\rmdefault}{\mddefault}{\updefault}{\color[rgb]{0,0,0}$9$}%
}}}}
\put(4201,-3811){\makebox(0,0)[b]{\smash{{\SetFigFont{12}{14.4}{\rmdefault}{\mddefault}{\updefault}{\color[rgb]{0,0,0}$10$}%
}}}}
\put(1051,-3586){\makebox(0,0)[rb]{\smash{{\SetFigFont{12}{14.4}{\rmdefault}{\mddefault}{\updefault}{\color[rgb]{0,0,0}$0$}%
}}}}
\put(1051,-3286){\makebox(0,0)[rb]{\smash{{\SetFigFont{12}{14.4}{\rmdefault}{\mddefault}{\updefault}{\color[rgb]{0,0,0}$2$}%
}}}}
\put(1051,-2986){\makebox(0,0)[rb]{\smash{{\SetFigFont{12}{14.4}{\rmdefault}{\mddefault}{\updefault}{\color[rgb]{0,0,0}$4$}%
}}}}
\put(1051,-2686){\makebox(0,0)[rb]{\smash{{\SetFigFont{12}{14.4}{\rmdefault}{\mddefault}{\updefault}{\color[rgb]{0,0,0}$6$}%
}}}}
\put(5551,-3286){\makebox(0,0)[rb]{\smash{{\SetFigFont{12}{14.4}{\rmdefault}{\mddefault}{\updefault}{\color[rgb]{0,0,0}$2$}%
}}}}
\put(5551,-2986){\makebox(0,0)[rb]{\smash{{\SetFigFont{12}{14.4}{\rmdefault}{\mddefault}{\updefault}{\color[rgb]{0,0,0}$4$}%
}}}}
\put(5551,-2686){\makebox(0,0)[rb]{\smash{{\SetFigFont{12}{14.4}{\rmdefault}{\mddefault}{\updefault}{\color[rgb]{0,0,0}$6$}%
}}}}
\put(1201,-5761){\makebox(0,0)[b]{\smash{{\SetFigFont{12}{14.4}{\rmdefault}{\mddefault}{\updefault}{\color[rgb]{0,0,0}$0$}%
}}}}
\put(1501,-5761){\makebox(0,0)[b]{\smash{{\SetFigFont{12}{14.4}{\rmdefault}{\mddefault}{\updefault}{\color[rgb]{0,0,0}$1$}%
}}}}
\put(1801,-5761){\makebox(0,0)[b]{\smash{{\SetFigFont{12}{14.4}{\rmdefault}{\mddefault}{\updefault}{\color[rgb]{0,0,0}$2$}%
}}}}
\put(2101,-5761){\makebox(0,0)[b]{\smash{{\SetFigFont{12}{14.4}{\rmdefault}{\mddefault}{\updefault}{\color[rgb]{0,0,0}$3$}%
}}}}
\put(2401,-5761){\makebox(0,0)[b]{\smash{{\SetFigFont{12}{14.4}{\rmdefault}{\mddefault}{\updefault}{\color[rgb]{0,0,0}$4$}%
}}}}
\put(2701,-5761){\makebox(0,0)[b]{\smash{{\SetFigFont{12}{14.4}{\rmdefault}{\mddefault}{\updefault}{\color[rgb]{0,0,0}$5$}%
}}}}
\put(3001,-5761){\makebox(0,0)[b]{\smash{{\SetFigFont{12}{14.4}{\rmdefault}{\mddefault}{\updefault}{\color[rgb]{0,0,0}$6$}%
}}}}
\put(3301,-5761){\makebox(0,0)[b]{\smash{{\SetFigFont{12}{14.4}{\rmdefault}{\mddefault}{\updefault}{\color[rgb]{0,0,0}$7$}%
}}}}
\put(3601,-5761){\makebox(0,0)[b]{\smash{{\SetFigFont{12}{14.4}{\rmdefault}{\mddefault}{\updefault}{\color[rgb]{0,0,0}$8$}%
}}}}
\put(3901,-5761){\makebox(0,0)[b]{\smash{{\SetFigFont{12}{14.4}{\rmdefault}{\mddefault}{\updefault}{\color[rgb]{0,0,0}$9$}%
}}}}
\put(4201,-5761){\makebox(0,0)[b]{\smash{{\SetFigFont{12}{14.4}{\rmdefault}{\mddefault}{\updefault}{\color[rgb]{0,0,0}$10$}%
}}}}
\put(1051,-5536){\makebox(0,0)[rb]{\smash{{\SetFigFont{12}{14.4}{\rmdefault}{\mddefault}{\updefault}{\color[rgb]{0,0,0}$0$}%
}}}}
\put(5701,-5761){\makebox(0,0)[b]{\smash{{\SetFigFont{12}{14.4}{\rmdefault}{\mddefault}{\updefault}{\color[rgb]{0,0,0}$0$}%
}}}}
\put(6001,-5761){\makebox(0,0)[b]{\smash{{\SetFigFont{12}{14.4}{\rmdefault}{\mddefault}{\updefault}{\color[rgb]{0,0,0}$1$}%
}}}}
\put(6301,-5761){\makebox(0,0)[b]{\smash{{\SetFigFont{12}{14.4}{\rmdefault}{\mddefault}{\updefault}{\color[rgb]{0,0,0}$2$}%
}}}}
\put(6601,-5761){\makebox(0,0)[b]{\smash{{\SetFigFont{12}{14.4}{\rmdefault}{\mddefault}{\updefault}{\color[rgb]{0,0,0}$3$}%
}}}}
\put(6901,-5761){\makebox(0,0)[b]{\smash{{\SetFigFont{12}{14.4}{\rmdefault}{\mddefault}{\updefault}{\color[rgb]{0,0,0}$4$}%
}}}}
\put(7201,-5761){\makebox(0,0)[b]{\smash{{\SetFigFont{12}{14.4}{\rmdefault}{\mddefault}{\updefault}{\color[rgb]{0,0,0}$5$}%
}}}}
\put(7501,-5761){\makebox(0,0)[b]{\smash{{\SetFigFont{12}{14.4}{\rmdefault}{\mddefault}{\updefault}{\color[rgb]{0,0,0}$6$}%
}}}}
\put(7801,-5761){\makebox(0,0)[b]{\smash{{\SetFigFont{12}{14.4}{\rmdefault}{\mddefault}{\updefault}{\color[rgb]{0,0,0}$7$}%
}}}}
\put(8101,-5761){\makebox(0,0)[b]{\smash{{\SetFigFont{12}{14.4}{\rmdefault}{\mddefault}{\updefault}{\color[rgb]{0,0,0}$8$}%
}}}}
\put(8401,-5761){\makebox(0,0)[b]{\smash{{\SetFigFont{12}{14.4}{\rmdefault}{\mddefault}{\updefault}{\color[rgb]{0,0,0}$9$}%
}}}}
\put(8701,-5761){\makebox(0,0)[b]{\smash{{\SetFigFont{12}{14.4}{\rmdefault}{\mddefault}{\updefault}{\color[rgb]{0,0,0}$10$}%
}}}}
\put(5551,-5536){\makebox(0,0)[rb]{\smash{{\SetFigFont{12}{14.4}{\rmdefault}{\mddefault}{\updefault}{\color[rgb]{0,0,0}$0$}%
}}}}
\put(1051,-5236){\makebox(0,0)[rb]{\smash{{\SetFigFont{12}{14.4}{\rmdefault}{\mddefault}{\updefault}{\color[rgb]{0,0,0}$2$}%
}}}}
\put(1051,-4936){\makebox(0,0)[rb]{\smash{{\SetFigFont{12}{14.4}{\rmdefault}{\mddefault}{\updefault}{\color[rgb]{0,0,0}$4$}%
}}}}
\put(1051,-4636){\makebox(0,0)[rb]{\smash{{\SetFigFont{12}{14.4}{\rmdefault}{\mddefault}{\updefault}{\color[rgb]{0,0,0}$6$}%
}}}}
\put(5551,-4636){\makebox(0,0)[rb]{\smash{{\SetFigFont{12}{14.4}{\rmdefault}{\mddefault}{\updefault}{\color[rgb]{0,0,0}$6$}%
}}}}
\put(5551,-4936){\makebox(0,0)[rb]{\smash{{\SetFigFont{12}{14.4}{\rmdefault}{\mddefault}{\updefault}{\color[rgb]{0,0,0}$4$}%
}}}}
\put(5551,-5236){\makebox(0,0)[rb]{\smash{{\SetFigFont{12}{14.4}{\rmdefault}{\mddefault}{\updefault}{\color[rgb]{0,0,0}$2$}%
}}}}
\put(5701,-3811){\makebox(0,0)[b]{\smash{{\SetFigFont{12}{14.4}{\rmdefault}{\mddefault}{\updefault}{\color[rgb]{0,0,0}$0$}%
}}}}
\put(2251,-2536){\makebox(0,0)[b]{\smash{{\SetFigFont{12}{14.4}{\rmdefault}{\mddefault}{\updefault}{\color[rgb]{0,0,0}req}%
}}}}
\put(6751,-2536){\makebox(0,0)[b]{\smash{{\SetFigFont{12}{14.4}{\rmdefault}{\mddefault}{\updefault}{\color[rgb]{0,0,0}req}%
}}}}
\put(7951,-2536){\makebox(0,0)[b]{\smash{{\SetFigFont{12}{14.4}{\rmdefault}{\mddefault}{\updefault}{\color[rgb]{1,0,0}gnt}%
}}}}
\put(2701,-4111){\makebox(0,0)[b]{\smash{{\SetFigFont{12}{14.4}{\rmdefault}{\mddefault}{\updefault}{\color[rgb]{0,0,0}(a)}%
}}}}
\put(7201,-4111){\makebox(0,0)[b]{\smash{{\SetFigFont{12}{14.4}{\rmdefault}{\mddefault}{\updefault}{\color[rgb]{0,0,0}(b)}%
}}}}
\put(2251,-4486){\makebox(0,0)[b]{\smash{{\SetFigFont{12}{14.4}{\rmdefault}{\mddefault}{\updefault}{\color[rgb]{0,0,0}req}%
}}}}
\put(3451,-4486){\makebox(0,0)[b]{\smash{{\SetFigFont{12}{14.4}{\rmdefault}{\mddefault}{\updefault}{\color[rgb]{1,0,0}gnt}%
}}}}
\put(6751,-4486){\makebox(0,0)[b]{\smash{{\SetFigFont{12}{14.4}{\rmdefault}{\mddefault}{\updefault}{\color[rgb]{0,0,0}req}%
}}}}
\put(7951,-4486){\makebox(0,0)[b]{\smash{{\SetFigFont{12}{14.4}{\rmdefault}{\mddefault}{\updefault}{\color[rgb]{1,0,0}gnt}%
}}}}
\put(2701,-6061){\makebox(0,0)[b]{\smash{{\SetFigFont{12}{14.4}{\rmdefault}{\mddefault}{\updefault}{\color[rgb]{0,0,0}(c)}%
}}}}
\put(7201,-6061){\makebox(0,0)[b]{\smash{{\SetFigFont{12}{14.4}{\rmdefault}{\mddefault}{\updefault}{\color[rgb]{0,0,0}(d)}%
}}}}
\put(3451,-2536){\makebox(0,0)[b]{\smash{{\SetFigFont{12}{14.4}{\rmdefault}{\mddefault}{\updefault}{\color[rgb]{1,0,0}gnt}%
}}}}
\end{picture}%

%% file: aircraft.pdf_t
\begin{picture}(0,0)%
\includegraphics{aircraft.pdf}%
\end{picture}%
\setlength{\unitlength}{3947sp}%
\begingroup\makeatletter\ifx\SetFigFont\undefined%
\gdef\SetFigFont#1#2#3#4#5{%
  \reset@font\fontsize{#1}{#2pt}%
  \fontfamily{#3}\fontseries{#4}\fontshape{#5}%
  \selectfont}%
\fi\endgroup%
\begin{picture}(8424,5124)(589,-6673)
\put(3001,-2686){\makebox(0,0)[b]{\smash{{\SetFigFont{12}{14.4}{\rmdefault}{\mddefault}{\updefault}{\color[rgb]{0,0,0}Left Elevator}%
}}}}
\end{picture}%